\documentclass[10pt,conference]{IEEEtran}
\usepackage{cite}
\usepackage{amsmath,amssymb}
\usepackage{graphicx}
\usepackage{booktabs}
\usepackage{listings}
\usepackage{url}
\usepackage{caption}
\usepackage{subcaption}
\usepackage{siunitx}
\usepackage{microtype}
\usepackage{balance}
\usepackage{color}
\usepackage{hyperref}
\usepackage{placeins}

\lstdefinestyle{mystyle}{
  basicstyle=\ttfamily\footnotesize,
  breaklines=true,
  numbers=left,
  numberstyle=\tiny\color{gray},
  frame=single,
  rulecolor=\color{gray!30},
  xleftmargin=8pt
}
\title{EMMap: A Systematic Framework for Spatial EMFI Mapping and Fault Classification on Microcontrollers\\}

\author{\IEEEauthorblockN{Gandham Sai Santhosh}
\IEEEauthorblockA{ee22b101@smail.iitm.ac.in \\
IIT Madras, India \\}
\and
\IEEEauthorblockN{Siddhartha Sanjay Naik}
\IEEEauthorblockA{ns25z175@smail.iitm.ac.in \\
IIT Madras, India \\}
\and
\IEEEauthorblockN{Ritwik Badola}
\IEEEauthorblockA{cs22s016@cse.iitm.ac.in \\
IIT Madras, India \\}
\and
\IEEEauthorblockN{Chester Rebeiro}
\IEEEauthorblockA{chester@cse.iitm.ac.in\\
IIT Madras, India \\}
}

\begin{document}
\maketitle

% ==============================================================
\begin{abstract}
Electromagnetic Fault Injection (EMFI) is a powerful technique for inducing bit flips and instruction-level perturbations on microcontrollers, yet existing literature lacks a unified methodology for systematically mapping spatial sensitivity and classifying resulting fault behaviors. Building on insights from O’Flynn\cite{OFlynn2020} and Kuhnapfel\cite{kuehnapfel2022} et al., we introduce a platform-agnostic framework for Spatial EMFI Mapping and Fault Classification, aimed at understanding how spatial probe position influences fault outcomes.

We present pilot experiments on three representative microcontroller targets including the Xtensa LX6 (ESP32) and two ChipWhisper boards not as definitive evaluations, but as illustrative demonstrations of how the proposed methodology can be applied in practice. These preliminary observations motivate a generalized and reproducible workflow that researchers can adopt when analyzing EMFI susceptibility across diverse embedded architectures.

\end{abstract}


% ==============================================================

\section{Introduction}

Electromagnetic Fault Injection (EMFI) has emerged as a practical and non-invasive technique for inducing transient faults in microcontrollers and system-on-chip (SoC) architectures. A carefully timed EM pulse can alter the execution path of a processor like skipping instructions, corrupting SRAM values, or bypassing security checks and thereby exposing subtle vulnerabilities in embedded systems. Prior studies have demonstrated this capability across architectures ranging from ARM microcontrollers to modern desktop-class processors. In particular, Kuhnapfel\cite{kuehnapfel2022} \textit{et al.} show that EMFI can reliably modify control flow, escape loops, and corrupt data when the injected pulse is aligned with narrowly defined execution windows, often synchronized with bus-level or GPIO events. Their work further highlights that environmental conditions, such as lowering the SoC supply voltage, substantially increase the likelihood of successful injection.

Similarly, O’Flynn's\cite{OFlynn2020} investigation into EMFI parameter space reveals that fault behavior is strongly influenced by physical characteristics of the injection setup: reversing coil polarity changes the dominant bit-flip type, adjusting probe height modulates fault magnitude, and scanning across the surface of a chip reveals regions with markedly different susceptibility. These insights collectively emphasize that EMFI outcomes are neither random nor uniformly distributed; rather, they arise from an interplay between spatial coupling, timing precision, and physical configuration.

Despite these advances, existing literature provides no unified methodology for systematically mapping \emph{where} on a chip faults can be induced and \emph{what type} of fault behavior arises from each spatial location. Many EMFI demonstrations remain highly platform-specific, with experimental details that do not generalize well across architectures. As a result, EMFI research often lacks the precision and repeatability needed for scientific rigor, and security evaluations remain dependent on ad-hoc, trial-and-error fault campaigns. Our motivation in this work is threefold. First, to move beyond coarse observations (``a glitch happened'') toward fault campaigns that consistently trigger specific behaviors such as instruction skips or bit-flips, enabling meaningful classification rather than stochastic success. Second, to identify which physical regions of a microcontroller are susceptible to EMFI and to document these vulnerabilities in a reproducible manner, transforming EMFI from a black-box technique into a spatially informed process. Third, to provide defenders and system designers with a structured approach to analyzing EMFI resilience, enabling targeted protections based on realistic, empirically grounded fault models.

\section{Proposed Framework for Spatial EMFI Mapping and Fault Classification}

The objective of this work is to develop a systematic and device-independent methodology for
\emph{spatial} characterization of EMFI susceptibility on microcontrollers, and for structured
classification of resulting fault behaviors. The framework is organized into three components: (1) spatial probing and parameter exploration, (2) architectural instrumentation and
fault observability, and (3) behavioral fault classification. Together, these components enable
repeatable experiments, comparison across platforms, and progressive refinement toward
targeted fault triggering.

\subsection{Spatial Probing and Parameter Exploration}

Spatial EMFI mapping begins with controlled variation of the probe position across the surface of
the target device. Independent of the commercial or custom EMFI toolchain used, the framework
assumes the following capabilities:

\begin{itemize}
    \item The die area (or package surface) is divided into a set of
    probe coordinates, typically arranged in a uniform grid. The resolution of the grid is not fixed;
    coarse scans are followed by finer scans around regions of interest.
    \item EMFI pulses must expose tunable parameters such as
    discharge voltage, pulse width, coil polarity, and probe height. Prior work shows that these
    parameters materially affect both magnitude and \emph{type} of resulting faults.
    \item The injection must be aligned with predictable program or
    hardware events. This includes GPIO toggles, bus transactions, or cycle markers within the
    firmware. Precise timing is essential, as EMFI success is strongly coupled to narrow execution
    windows.
\end{itemize}

The outcome of this stage is a multi-dimensional parameter space describing how spatial position
and pulse parameters relate to observable deviations, forming the basis for subsequent fault
classification.

\subsection{Architectural Instrumentation and Fault Observability}

Since internal microarchitectural state is generally inaccessible on production MCUs, the framework
requires explicit instrumentation at the firmware level to detect and differentiate fault outcomes.
To maintain generality across architectures, we adopt simple but expressive observability
mechanisms:

\begin{itemize}
    \item Tight loops, counters, or instruction sequences
    whose nominal behavior is known and easily checked for skips or early exits.
    \item Small blocks of SRAM with fixed patterns or CRCs to detect
    data corruption without requiring full memory dumps.
    \item UART, GPIO signals, or debug interfaces that emit
    parseable markers for automated logging.
\end{itemize}

These allow the framework to detect instruction-level disturbances,
transient SRAM corruptions, or system-level anomalies (e.g., hangs) independent of the specific
microcontroller used.

\subsection{Behavioral Fault Classification}

The final component of the framework organizes observed deviations into a hierarchy of fault classes.
This is crucial for both scientific rigor and security evaluation. We distinguish three primary classes:

\begin{enumerate}
    \item \textbf{Control-flow faults:} instruction skips, altered loop bounds, premature exits, or incorrect
    branch outcomes.
    \item \textbf{Data corruption faults:} bit-flips in sentinel memory regions, register perturbations,
    or multi-byte corruption patterns.
    \item \textbf{System-level faults:} hangs, resets, peripheral misbehavior, or error-flag activation.
\end{enumerate}

Faults collected across multiple spatial coordinates and pulse configurations are used to build a
\emph{spatial susceptibility map} of the device. These maps serve as empirical approximations of the
underlying physical coupling between the EM field and microarchitectural structures.

\begin{figure}[t]
    \centering
    \includegraphics[width=0.8\linewidth]{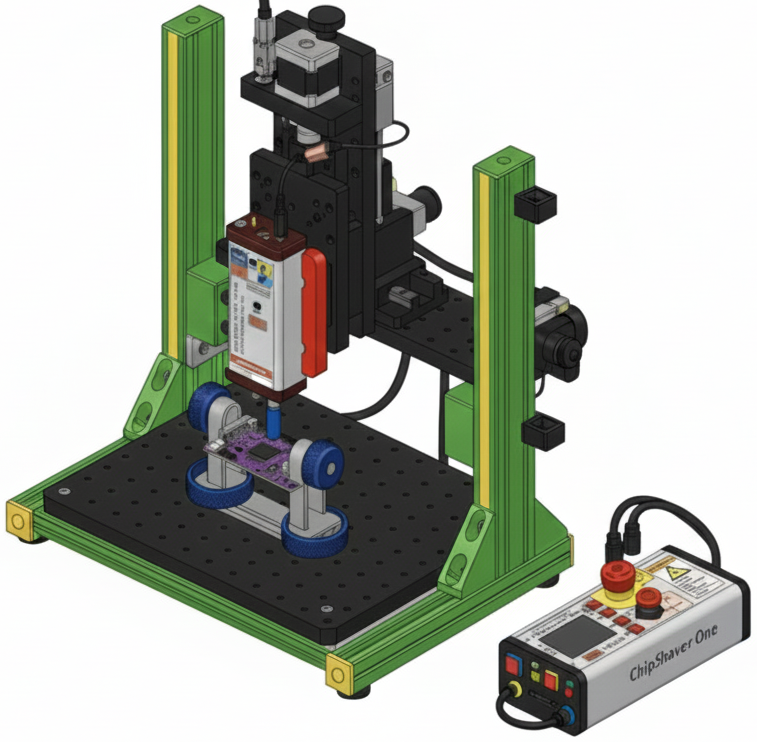}
    \caption{CAD-style illustration of the EMFI setup used in this work, including the XYZ motion stage, ChipSHOUTER pulse generator, EMFI probe, 3D-printed target holder, and ESP32 target board.}
    \label{fig:emfi_setup}
\end{figure}

This section summarizes how the proposed framework was instantiated on three representative
microcontroller-class platforms. The goal of these experiments is not to provide exhaustive device
characterization, but to demonstrate how each stage of the framework which is spatial probing,
instrumentation, and fault classification can be operationalized in practice.

\subsection{Target Platforms}

To exercise the framework across heterogeneous architectures, we selected three platforms with
distinct levels of visibility into internal state:

\begin{itemize}
    \item \textbf{Tensilica Xtensa LX6 (ESP32):} provides full debug access, enabling register snapshots
    and breakpoint-controlled execution. This allows fine-grained observation of control-flow and
    architectural state under EMFI.
    \item \textbf{ChipWhisperer CW521 (external SRAM target):} contains an off-chip memory region
    isolated from any processor. This enables direct observation of bit-level memory corruption without
    interference from pipelines or caches.
    \item \textbf{ChipWhisperer CW322 (STM32F3 MCU):} integrates CPU, SRAM, and peripherals on-die,
    representing a realistic embedded target where only high-level program behavior is observable.
\end{itemize}

Together, these platforms allow demonstration of the framework across scenarios ranging from
high observability (ESP32) to highly constrained embedded conditions (CW322).

\subsection{Spatial Scanning Configuration}

Each platform was scanned using a uniform grid of spatial probe positions, with resolution chosen
according to practical limits of probe movement and target form factor. Coarse grids were used to
identify regions of elevated susceptibility, while finer grids were applied selectively around these
regions for refinement. For each coordinate, multiple EMFI pulses were injected under a fixed set
of pulse parameters to capture statistical behavior.  

Consistent with prior reports, we observed that variation in coil polarity, probe height, and discharge
parameters meaningfully affected the probability and type of observed deviations. These variations
are treated within the framework as part of the parameter-exploration stage rather than as fixed
properties of any specific platform.

\begin{figure*}[!t]
  \centering
  \begin{subfigure}[b]{0.4\textwidth}
    \includegraphics[width=\linewidth]{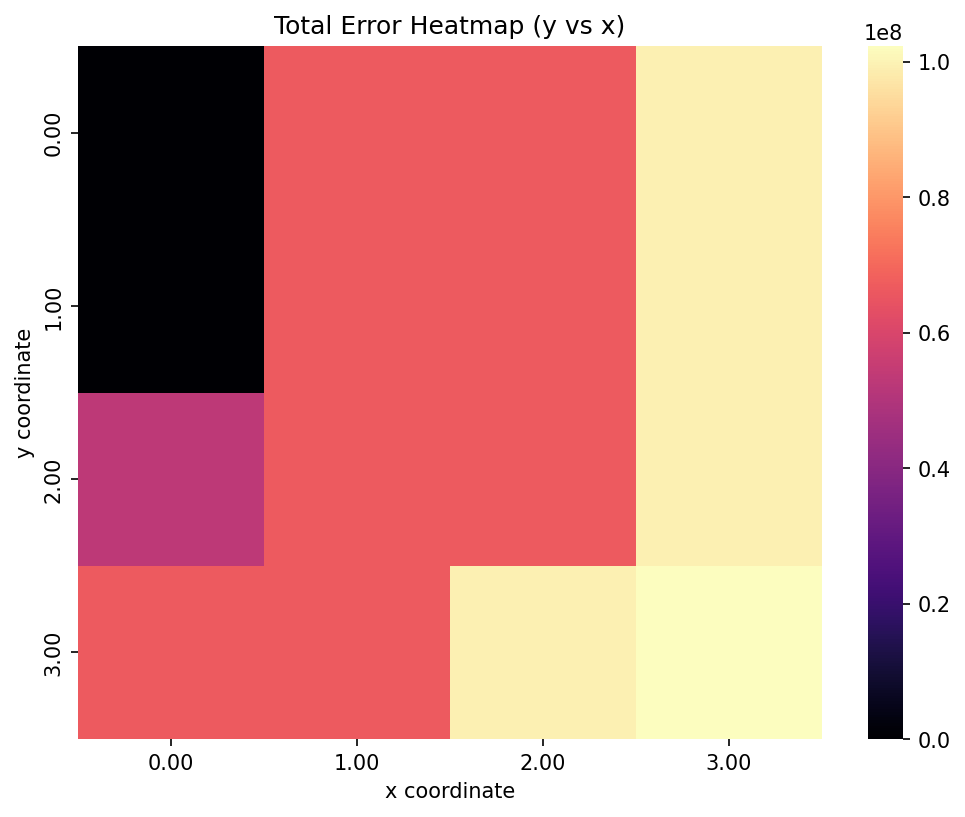}
    \caption{2-D error heatmap (per-coordinate error counts).}
    \label{fig:heatmap}
  \end{subfigure}
  \hfill
  \begin{subfigure}[b]{0.4\textwidth}
    \includegraphics[width=\linewidth]{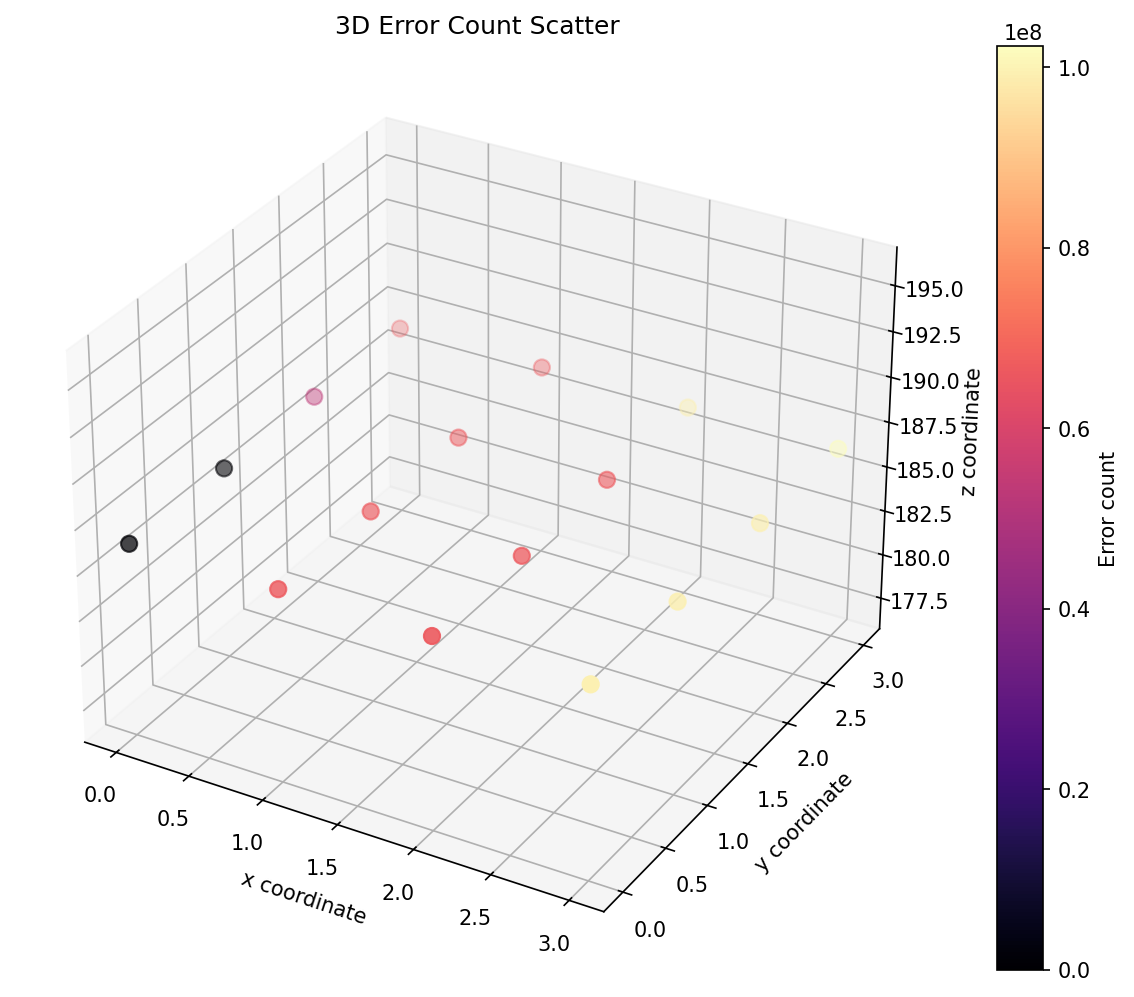}
    \caption{3-D scatter (x, y, error count / repeat index).}
    \label{fig:scatter}
  \end{subfigure}
  \caption{Representative CW521 outputs: heatmap and 3D error scatter from pilot runs.}
  \label{fig:results}
\end{figure*}

\subsection{Firmware Instrumentation for Observability}

Following the framework, each platform was instrumented with minimal but expressive observability
mechanisms:

\begin{itemize}
    \item \textbf{Xtensa LX6:} breakpoints were inserted within a reference program path.
    At each halt, register states (PC, general-purpose registers, and local disassembly) were captured
    automatically. This enabled distinguishing transient register disturbances from stable architectural
    behavior.
    \item \textbf{CW521:} memory patterns were written to the external SRAM, and full read-back
    comparisons were performed following injection. This allowed direct enumeration of bit-flip
    locations and magnitudes, serving as an illustrative example of physical data corruption mapping.
    \item \textbf{CW322:} firmware-level sentinel memory blocks with CRC checks were incorporated,
    along with deterministic loop counters that produced predictable control-flow markers.
    The resulting UART or GPIO outputs provided a lightweight means of detecting instruction skips,
    data corruption, or hangs.
\end{itemize}

These instrumentation mechanisms were selected to align with the framework requirements rather
than to optimize results for any specific device.

\subsection{Illustrative Observations}

Across the three platforms, we observed representative instances of the fault classes defined in the
framework:

\begin{itemize}
    \item ESP32 occasionally halted with altered register states, and the STM32-based CW322 exhibited skipped loop iterations or premature exits.
    \item CW521 external SRAM showed localized bit flips under certain spatial and pulse configurations, consistent with expectations for an isolated memory device.
    \item Resets, hangs, or desynchronized UART outputs appeared on the MCU-based platforms under specific timing alignments.
\end{itemize}
\vspace{1mm}
To illustrate how spatial EMFI behavior manifests in practice, Fig. 2 summarizes representative outputs from the CW521 external-SRAM platform. The 2-D heatmap in Fig. 2(a) aggregates the total number of observed bit-flip events per probe coordinate across repeated injections. The resulting spatial gradient serves as a coarse susceptibility map: darker regions correspond to coordinates where no corruption was detected, whereas lighter regions highlight probe locations that consistently induced high-magnitude disturbances. Because CW521 exposes only memory-level effects, this visualization provides a direct spatial signature of physical coupling between the EM pulse and the SRAM array.

\vspace{1mm}
Fig. 2(b) complements this view by plotting a 3-D scatter of individual fault events, where each point encodes the (x, y) probe coordinate and the associated error count for a single trial. This representation makes the probabilistic nature of EMFI readily visible. Points with similar spatial coordinates exhibit varying error magnitudes, reinforcing that even at a fixed location, the induced disturbance amplitude depends on micro-scale EM coupling, stochastic charge redistribution, and pulse-to-pulse variation. Together, these plots demonstrate that spatial position is a primary determinant of corruption density on the CW521 device and that repeated injections are necessary to correctly estimate susceptibility at any given coordinate.

\vspace{1mm}
These observations are preliminary and are included strictly to illustrate the application of the
framework. No claims are made regarding completeness, reproducibility across all conditions, or
device-specific vulnerability conclusions.

\subsection{Role of These Demonstrations}

The primary role of these demonstrations is to validate that the proposed framework can be applied
to diverse microcontroller environments with varying observability constraints. The experiments
provide qualitative evidence that:

\begin{enumerate}
    \item spatial probe position has measurable impact on fault behavior,
    \item simple firmware instrumentation is sufficient for high-level fault classification,
    \item and the framework can organize observed behavior into coherent, cross-platform fault classes.
\end{enumerate}

These findings motivate the more general discussion and systematization presented in the following
section.

\section{Discussion: Systematizing Spatial EMFI Behavior}

The preliminary demonstrations across the three platforms highlight broader principles that
underlie spatial EMFI behavior. Although each device differs in architecture, memory hierarchy,
and observability, several consistent themes emerge when the framework is applied. This section
systematizes these themes and articulates general insights that inform how spatial mapping and
fault classification should be interpreted by practitioners.

% \FloatBarrier
\begin{figure*}[!t]
  \centering
  \begin{subfigure}[b]{0.48\textwidth}
    \includegraphics[width=\linewidth]{pipeline.png}
    \caption{Pipeline for spatial EMFI experimentation showing the probe setup, 
    JTAG interface, and automated host-side tooling.}
    \label{fig:pipeline}
  \end{subfigure}
  \hfill
  \begin{subfigure}[b]{0.48\textwidth}
    \includegraphics[width=\linewidth]{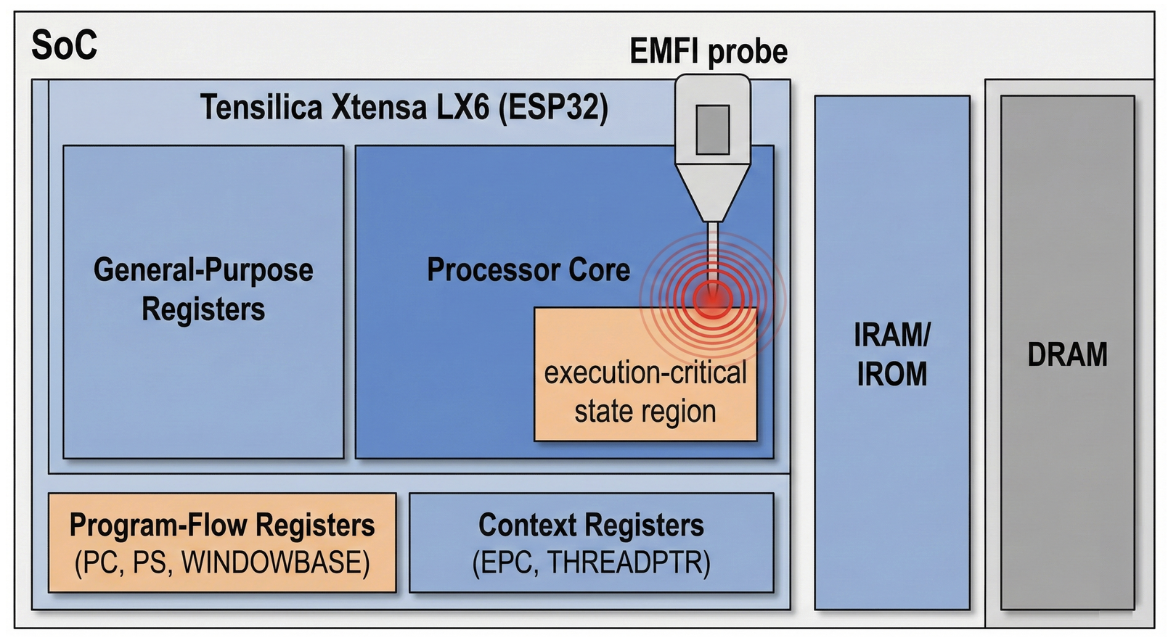}
    \caption{Conceptual diagram of EMFI interaction with the Xtensa LX6 core, 
    highlighting the execution-critical state region.}
    \label{fig:xtensa}
  \end{subfigure}

  \caption{Visual Overview of Fault Injection Analysis on Xtensa LX6 (ESP32)}
  \label{fig:results}
\end{figure*}

\subsection{Spatial Sensitivity as a Fundamental Property}

Prior work has clearly shown that EMFI susceptibility is not uniform across a chip surface.
O’Flynn’s results demonstrate that even millimeter-scale movements of the probe can
meaningfully alter both the likelihood and the type of induced faults. Similarly, Kuhnapfel
\textit{et al.} observe that specific subregions of a large SoC exhibit markedly different behavior under
identical pulse conditions.

Our demonstrations reinforce this principle: spatial position consistently influences whether the
observed faults fall into control-flow, data-corruption, or system-level categories. Even when
internal mapping (e.g., die layout or memory banks) is unknown, spatially organized sampling
reveals meaningful gradients in susceptibility. The framework therefore treats spatial mapping as
a first-class component, rather than a secondary tuning step.

\subsection{Timing Precision Remains a Critical Axis}

Across all platforms, the success of injection attempts depended strongly on temporal alignment
between the EM pulse and specific execution windows. This aligns with prior observations that
only narrow timing margins around instruction fetch, memory read, or pipeline transitions produce
reliable disturbances.

While the demonstrations in this work employ basic timing markers (e.g., GPIO toggles or loop
cycles), the framework is compatible with higher-precision tools such as FPGA-based delay
generators or bus-triggered injection. Importantly, the necessity for temporal precision is treated
as architecture-independent: any microcontroller with deterministic execution sequences can serve
as a substrate for repeatable timing-aligned fault campaigns.

\subsection{Physical Parameters as Fault-Mode Modulators}

The EMFI parameter space is multidimensional. Coil polarity, probe height, discharge voltage,
pulse width, and local EM coupling all have nontrivial influence on fault outcomes. O’Flynn's
observations that reversing coil polarity shifts the dominant bit-flip direction and that reducing
probe height increases fault magnitude were reflected qualitatively across our experimental
platforms.

Within the framework, these parameters are not treated as fixed experimental constants. Instead,
they serve as \emph{fault-mode modulators} that help distinguish whether observed phenomena arise
from localized memory disturbance, timing violations, or system-level perturbations. This
classification supports reproducibility without requiring exact replication of hardware conditions.

\subsection{Classification Enables Structured Interpretation}

The framework's three-tier classification scheme: control-flow faults, data-corruption faults, and
system-level faults proved expressive enough to organize all observed deviations, despite the
differences among platforms. This validates that the taxonomy is practical and sufficiently
agnostic to architectural details.

The distinction is especially important for MCU-class devices where internal state visibility is
limited. For example, a skipped loop iteration and a CRC mismatch may appear superficially
similar but arise from fundamentally different microarchitectural effects. By categorizing these
faults separately, the framework enables more meaningful cross-platform comparisons and
supports clearer reasoning about underlying mechanisms.

\subsection{Toward a Generalized Methodology}

The combination of spatial sensitivity, temporal precision, and physical modulation suggests that
a generalized EMFI mapping methodology should begin with coarse spatial scans to identify broad susceptibility regions, employ lightweight instrumentation to distinguish fault classes, iteratively refine both spatial resolution and pulse parameters, and interpret observations through the lens of architecture-independent principles rather than device-specific heuristics.

\subsection{Limitations and Scope}

As with all preliminary studies, the demonstrations presented here serve primarily to validate the
\emph{applicability} of the framework rather than to characterize any particular device. The results
are intentionally conservative: no claims are made regarding completeness, exact spatial
correspondence to die structures, or vulnerability of the tested platforms. Instead, the value lies in
showing that the framework can structure experimentation in a principled and architecture-agnostic
manner.

These limitations outline a natural direction for future work, where more precise timing
instrumentation, physical die imaging, or higher-resolution spatial scans may yield deeper
insights into the physical origins of observed behaviors.

\section{Practical Guidelines for Applying the Framework}

While the previous sections define the conceptual structure of the framework, effective use in real
EMFI campaigns requires concrete experimental practices. The following guidelines consolidate
lessons from prior literature and our illustrative demonstrations into a set of practical
recommendations for researchers and evaluators.

\subsection{Timing Alignment and Triggering}

EMFI effectiveness is strongly dependent on alignment with specific execution windows. To
maximize interpretability:

\begin{itemize}
    \item \textbf{Use deterministic firmware markers:} GPIO toggles, loop iteration points, or UART tokens
    provide coarse alignment without relying on internal debug hardware.
    \item \textbf{Favor narrow windows:} even simple time-offset sweeps can reveal windows of heightened
    vulnerability, consistent with prior observations that many faults require precise synchronization.
    \item \textbf{Record timing offsets explicitly:} timestamps, trigger delays, and event markers should be
    captured for each injection to support later analysis.
\end{itemize}

Accurate timing metadata allows researchers to correlate spatial and temporal effects more reliably.

\subsection{Firmware Instrumentation for Observability}

Because most microcontrollers do not expose internal state, firmware instrumentation is crucial:

\begin{itemize}
    \item \textbf{Employ sentinel memory regions:} checksums or fixed patterns provide fast, lightweight
    detection of data corruption.
    \item \textbf{Use predictable control-flow constructs:} tight loops, constant-iteration counters, and
    simple arithmetic blocks make instruction skips immediately observable.
    \item \textbf{Generate structured output:} unambiguous UART or GPIO messages (e.g.,
    \texttt{CF\_SKIP}, \texttt{CRC\_ERR}) facilitate automated logging and classification.
\end{itemize}

\subsection{Fault Classification Workflow}

To maintain consistent interpretation across devices:

\begin{enumerate}
    \item Determine whether the deviation corresponds to control-flow,
    data corruption, or system-level behavior.
    \item Measure loop iteration counts, bit-flip density, or reset
    patterns to gain deeper insight.
    \item EMFI is probabilistic; classification should be based
    on repeated trials under identical parameters.
\end{enumerate}

\subsection{Reproducibility and Documentation Practices}

To support scientific rigor, the following practices are recommended:

\begin{itemize}
    \item Probe geometry, pulse parameters, timing offsets, and firmware
    versions should be logged for every experiment.
    \item Structured logs and repeatable scan sequences
    reduce manual error.
    \item Supply voltage, temperature, and noise sources can influence outcomes and should be documented when relevant.
\end{itemize}

\subsection{Extensibility Toward Advanced Analysis}

The framework is compatible with more sophisticated techniques when needed:

\begin{itemize}
    \item \textbf{High-resolution timing:} FPGA- or logic-analyzer-based triggering can reveal finer-grained
    vulnerability windows.
    \item \textbf{Die-level spatial correlation:} microscopy or decapsulation allows aligning susceptibility
    maps with physical structures, as demonstrated in prior work on large SoCs.
    \item \textbf{Automated parameter search:} scripted sweeps of spatial, temporal, or pulse parameters
    can accelerate exploration of the fault space.
\end{itemize}

\vspace{1mm}
Overall, these guidelines operationalize the framework in a manner that is reproducible, device-agnostic,
and compatible with EMFI infrastructure.

\section{Conclusion}

This work presents a systematic, platform-agnostic framework for spatial EMFI mapping and fault
classification on microcontrollers. Motivated by the need for precision, repeatability, and scientific
rigor in EMFI experimentation, the framework unifies spatial probing, firmware-level
instrumentation, and structured fault classification into a coherent methodology. Rather than
focusing on device-specific vulnerabilities, the approach emphasizes general principles that hold
across architectures: spatial susceptibility gradients, narrow timing windows for effective
injection, and the modulatory influence of physical parameters such as coil polarity and probe
height.

Pilot demonstrations on the Xtensa LX6 (ESP32) and two ChipWhisperer-based educational
targets illustrate how the methodology can be instantiated under varying levels of observability.
These experiments serve not as comprehensive evaluations but as practical examples of how the
framework organizes EMFI behavior into interpretable categories and highlights the interplay
between spatial, temporal, and physical dimensions.

By formalizing this workflow, the framework provides a foundation for reproducible EMFI
research and offers a structured path for analysts seeking to characterize the fault-injection
susceptibility of embedded systems. It also supports a more transparent and disciplined approach
to security evaluation, enabling designers to reason about EMFI resilience using empirically
grounded fault models.

Future extensions may incorporate high-resolution timing control, die-level imaging for spatial
correlation, or automated multi-parameter exploration. Nonetheless, the core methodology
presented here establishes a generalizable and extensible basis for advancing spatial EMFI
characterization across diverse microcontroller platforms.


\begin{thebibliography}{99}

\bibitem{OFlynn2020}
C.~O'Flynn, ``NAEAN0011: Electromagnetic Fault Injection (EMFI) for Automotive
Safety \& Security Testing with ChipSHOUTER{\textregistered},'' NewAE
Technology, Tech. Rep., 2020.
Available: \url{https://media.newae.com/appnotes/NAE0011_Whitepaper_EMFI_For_Automotive_Safety_Security_Testing.pdf}.

\bibitem{kuehnapfel2022}
N. Kühnapfel, R. Buhren, H.-N. Jacob, T. Krachenfels, C. Werling, J.-P. Seifert, 
``EM-Fault It Yourself: Building a Replicable EMFI Setup for Desktop and Server Hardware,'' 
PAINE / arXiv preprint, Sept. 2022.  
Available: \url{https://arxiv.org/abs/2209.09835}.

\bibitem{chipshouter_manual}
NewAE Technology Inc., 
``CHIPSHOUTER® USER MANUAL,'' Feb 25, 2021.  
Available: \url{https://media.newae.com/manuals/ChipSHOUTER_PRESS_1.3.pdf}.

\bibitem{badfet2017}
A. Cui and R. Housley, 
``BADFET: Defeating Modern Secure Boot Using Second-Order Pulsed Electromagnetic Fault Injection,'' 
WOOT (USENIX Workshop on Offensive Technologies), 2017.  
Available: \url{https://www.usenix.org/system/files/conference/woot17/woot17-paper-cui.pdf}.

\bibitem{moro2014}
N. Moro, A. Dehbaoui, K. Heydemann, B. Robisson, E. Encrenaz, 
``Electromagnetic Fault Injection: Towards a Fault Model on a 32-bit Microcontroller,'' 
arXiv / FDTC papers, 2014.  
Available: \url{https://arxiv.org/pdf/1402.6421.pdf}.

\bibitem{trouchkine2021}
T. Trouchkine, G. Bouffard, J. Clédière, 
``EM Fault Model Characterization on SoCs: From Different Architectures to the Same Fault Model,'' 
FDTC (Workshop on Fault Detection and Tolerance in Cryptography), 2021.  
Available: \url{https://thomas.trouchkine.com/assets/pdf/fdtc_2021.pdf}.

\bibitem{liao2019}
H. Liao and C. Gebotys, 
``Methodology for EM Fault Injection: Charge-based Fault Model,'' 
DATE (Design, Automation and Test in Europe), 2019.  
Available: \url{https://past.date-conference.com/proceedings-archive/2019/DATA/401.pdf}.

\bibitem{abdellatif2020_silicontoaster}
K. M. Abdellatif and O. Hériveaux, 
``SiliconToaster: A Cheap and Programmable EM Injector for Extracting Secrets,'' 
FDTC Workshop (FDTC/2020), 2020.  
Available: \url{https://fdtc.deib.polimi.it/FDTC20/shared/Abdellatif.pdf}.

\bibitem{balasch2017}
J. Balasch, D. Arumí, S. Manich, 
``Design and validation of a platform for electromagnetic fault injection,'' 
DCIS (Design of Circuits and Integrated Systems), 2017.  
Available (PDF): \url{https://www.researchgate.net/publication/323823120_Design_and_validation_of_a_platform_for_electromagnetic_fault_injection}.

\bibitem{toulemont2020}
J. Toulemont, N. Ouldei-Tebina, J.-M. Galliére, P. Nouet, E. Bourbao, P. Maurine, 
``A Simple Protocol to Compare EMFI Platforms,'' 
IACR Cryptology ePrint Archive, 2020 (eprint).  
Available: \url{https://eprint.iacr.org/2020/1277.pdf}.

\bibitem{buhren2021}
R. Buhren, H.-N. Jacob, T. Krachenfels, J.-P. Seifert, 
``One Glitch to Rule Them All: Fault Injection Attacks Against AMD's Secure Encrypted Virtualization,'' 
ACM CCS (preprint / arXiv), 2021.  
Available: \url{https://arxiv.org/abs/2108.04575}.

\bibitem{kim2014_rowhammer}
Y. Kim, R. Daly, J. Kim, C. Fallin, J. H. Lee, D. Lee, C. Wilkerson, K. Lai, O. Mutlu, 
``Flipping Bits in Memory Without Accessing Them: An Experimental Study of DRAM Disturbance Errors,'' 
ISCA, 2014.  
Available: \url{https://people.inf.ethz.ch/omutlu/pub/rowhammer_isca2014.pdf}.

\bibitem{dutertre2019}
J.-M. Dutertre, T. Riom, O. Potin, J.-B. Rigaud, 
``Experimental Analysis of the Laser-Induced Instruction Skip Fault Model,'' 
NordSec / Microelectronics Reliability (2019/2021 versions), 2019.  
Available (paper / slides): \url{https://hal.science/hal-02420090/document}.

\bibitem{delarea2022}
S. Delarea and Y. Oren, 
``Practical, Low-Cost Fault Injection Attacks on Personal Smart Devices,'' 
Applied Sciences, vol. 12, no. 1, 2022.  
Available: \url{https://www.mdpi.com/2076-3417/12/1/417}.

\bibitem{silicon_toaster_repo}
SiliconToaster / authors, 
``SiliconToaster project materials (FDTC paper + code),'' 2020.  
Available: \url{https://fdtc.deib.polimi.it/FDTC20/shared/Abdellatif.pdf}.

\bibitem{riscure_emps_datasheet}
Riscure, 
``Riscure EM Probe Station 5 (datasheet) / EM-FI Transient Probe,'' product datasheet.  
Available: \url{https://getquote.riscure.com/picdb/filedb/3792/EMPS%205%20datasheet.pdf}.

\bibitem{troubchkine2021_fdtc}
T. Trouchkine, G. Bouffard, J. Clédière, 
``EM Fault Model Characterization on SoCs: From Different Architectures to the Same Fault Model,'' 
FDTC 2021 (paper).  
Available: \url{https://thomas.trouchkine.com/assets/pdf/fdtc_2021.pdf}.

\bibitem{liao2019_date}
H. Liao and C. Gebotys, 
``Methodology for EM Fault Injection: Charge-based Fault Model,'' 
DATE 2019 proceedings (paper), 2019.  
Available: \url{https://past.date-conference.com/proceedings-archive/2019/DATA/401.pdf}.

\bibitem{voltpillager_usenix2021}
Z. Chen, G. Vasilakis, K. Murdock, E. Dean, D. Oswald, F. D. Garcia, 
``VoltPillager: Hardware-based fault injection attacks against Intel SGX Enclaves using the SVID voltage scaling interface,'' 
USENIX Security 2021 (paper).  
Available: \url{https://www.usenix.org/conference/usenixsecurity21/presentation/chen-zitai}.

\bibitem{v0ltpwn_usenix2020}
Z. Kenjar, T. Frassetto, D. Gens, M. Franz, A.-R. Sadeghi, 
``V0LTpwn: Attacking x86 Processor Integrity from Software,'' 
USENIX Security 2020.  
Available: \url{https://www.usenix.org/conference/usenixsecurity20/presentation/kenjar}.

\bibitem{voltjockey_2020}
P. Qiu, D. Wang, Y. Lyu, R. Tian, C. Wang, G. Qu, 
``VoltJockey: A New Dynamic Voltage Scaling based Fault Injection Attack on Intel SGX,'' 
IEEE Transactions on CAD (2020).  
Available: \url{https://ieeexplore.ieee.org/document/9083976}.

\bibitem{chipshouter_github}
NewAE Technology, 
``ChipSHOUTER (GitHub repository),'' NewAE / ChipSHOUTER project page.  
Available: \url{https://github.com/newaetech/ChipSHOUTER}.

\end{thebibliography}
\end{document}